\documentclass[10pt]{article}
\usepackage{amsmath,amssymb,amsfonts}
\usepackage{calc}
\usepackage{hyperref} 
\usepackage{graphicx}
\usepackage{color}
\usepackage{listings}
\usepackage{amsmath,amssymb,amsthm,bbm}
\usepackage{marginnote}

\usepackage{tikzpagenodes}
\usepackage{fourier-orns}
\usepackage{lipsum}
\usepackage[T1]{fontenc}
\usepackage{geometry}
\usepackage{titlesec}  
\usepackage{xcolor}
\usepackage{amsmath,amssymb,amsfonts,mathrsfs,datetime}
\usepackage{tikz}
\usepackage{graphics}
\usetikzlibrary{calc,decorations.markings}
\usetikzlibrary{datavisualization.formats.functions}

\usepackage{tcolorbox}
\usepackage{amsthm}
\usepackage{graphicx}
\usetikzlibrary{calc,decorations.markings}
\usetikzlibrary{datavisualization.formats.functions}
\def\beq{\begin{eqnarray}}
\def\eeq{\end{eqnarray}}
\usepackage{accents}
\newcommand{\dbtilde}[1]{\accentset{\approx}{#1}}

\newtheorem{theorem}{Theorem}

\newtheorem{rule-of-thumb}[theorem]{Definition} 
\definecolor{purple}{rgb}{0.7,0,0.7}
\definecolor{acquamarine}{rgb}{0.,0.4,0.5}

\begin{document}

\title{The heterogeneous helicoseir }
\author{Paolo Amore \\
{\small Facultad de Ciencias, CUICBAS, Universidad de Colima},\\
{\small Bernal D\'{i}az del Castillo 340, Colima, Colima, Mexico} \\
{\small paolo@ucol.mx} \\
John P. Boyd \\
{\small Department of Atmospheric, Oceanic \& Space Science}\\
{\small University of Michigan, 2455 Hayward Avenue, Ann Arbor, MI 48109, United States}\\
{\small jpboyd@umich.edu } \\
Abigail M\'arquez   \\
{\small Facultad de Ciencias, Universidad de Colima},\\
{\small Bernal D\'{i}az del Castillo 340, Colima, Colima, Mexico} \\
{\small vmarquez2@ucol.mx } }

\maketitle

\begin{abstract}
We study the rotations of a heavy string (helicoseir) about a vertical axis with one free endpoint and with arbitrary density, under the action of the gravitational force. We show that the problem can be transformed into a nonlinear eigenvalue equation, as in the  uniform case. The eigenmodes of this equation represent equilibrium configurations of the rotating string in which the shape of the string doesn't change with time. As previously proved by Kolodner for the homogenous case, the occurrence of new modes of the nonlinear equation  is tied  to the spectrum of the corresponding linear equation. We have been able to generalize this result to a class of densities $\rho(s) = \gamma (1-s)^{\gamma-1}$, which includes the homogenous string as a special case ($\gamma=1$). 

We also show that the solutions to the nonlinear eigenvalue equation (NLE) for an arbitrary density are orthogonal and that 
a solution of this equation with a given number of nodes contains solutions of a different helicoseir, with a smaller number of nodes. Both properties hold also for the homogeneous case and had not been established before. 
\end{abstract}

\maketitle



\section{Introduction}
\label{sec:intro}

If a heavy string is set to rotate uniformly about a vertical axis, it can display stable configurations in which its shape is not 
changing with time, as a result of the balance between the gravitational and the centrifugal forces acting on it. 
As the angular velocity is increased, the string can assume more complicated shapes, with an increasing number of nodes, namely points in which the string crosses the axis of rotation. This problem has been first considered by Wilson in ref.~\cite{Wilson08}, where he found that  $\omega^2 l/g > 3/2$ in order to obtain non vertical configurations of equilibrium~\footnote{Edwin Bidwell Wilson, who wrote the mathematical treatment of the helicoseir more than a century ago, had a remarkable career in mathematics, physics and statistics as detailed in \cite{Carvajalino18} and \cite{HunsakerMacLane73}.}; much later (1955), Kolodner proved for the uniform string that for $\lambda_n < \omega^2 l/g \leq \lambda_{n+1}$, with $\lambda_n \equiv \sigma_n^2/4$ ($\sigma_n$ is the nth zero of the Bessel function of order $0$) the nonlinear eigenvalue equation has exactly $n$ non-trivial solutions (theorem 1 of ref.~\cite{Kolodner55}).
This theorem implies that the string will depart from the vertical for $\omega^2 l/g > \sigma_1^2/4 \approx 1.4458$, thus refining the qualitative estimate of Wilson.
The special values $\lambda_n$ found by Kolodner are the eigenvalues of the linearized problem which can be obtained from the nonlinear equation, by considering only small deviations from the axis of rotation; Kolodner proved that the eigenvalues of the linear problem (which form a  discrete spectrum) are thresholds for the appearance of new modes in the nonlinear problem. Antman~\cite{Antman80} has extended Kolodner analysis to the case of heavy elastic strings.

In more recent times the problem has been revisited by several authors: in particular Silverman, Strange and Lipscombe in ref.~\cite{Silverman98} studied numerically the solutions corresponding to large amplitude (fully nonlinear problem) and compared them with the corresponding solutions of the linearized problem; they also performed an experiment using a uniform keychain which was set to rotate at different angular velocities. Coomer et al.~\cite{Coomer01} also studied the nonlinear eigenvalue equation associated to this problem and the corresponding experimental whirling modes.

Dmitrochenko and collaborators in refs.~\cite{Dmitrochenko06a,Dmitrochenko06b} discussed the stability of the equilibrium configurations; the unpublished manuscript by Carter~\cite{Carter09} contains a detailed discussion of the helicoseir problem, while the recent paper by Denny~\cite{Denny20} discusses solutions to the nonlinear problem at high rotation rate; finally Pham and Pham in \cite{Pham17} study the problem of robotic manipulation of a rotating chain.

In this paper we consider the more general case of a string with an arbitrary density and derive the corresponding nonlinear eigenvalue equation; the results obtained by Kolodner for the uniform string are generalized to the case 
of a string with density $\rho(s) = \gamma (1-s)^{\gamma-1}$ ($\gamma=1$ correponds to the uniform string); also in this case we find that the eigenvalues of the linearized problem act as thresholds for the appearance of new nonlinear modes.
In addition to this, we have established some general properties of the nonlinear solutions, which had not been noticed before: as a matter of fact we prove that the solutions to the nonlinear eigenvalue equation obey a \emph{nonlinear} orthogonality condition; 
additionally  we show that it is possible to obtain new solutions to the nonlinear eigenvalue equation, starting from a particular solution  with $n >1$ internal nodes (something that we jokingly refer to as a "matrioska effect").

The paper is organized as follows: in section \ref{sec:hh} we derive the nonlinear eigenvalue equation for the heterogeneous problem
and discuss some of its properties; in section \ref{sec:eq} we  discuss the conditions of equilibrium and estimate the minimal angular 
velocity at which a non--vertical equilibrium can be reached (for a homogeneous string this reduces to the result of Wilson \cite{Wilson08}); in section \ref{sec:rs} we prove that the solutions to the nonlinear eigenvalue equation with more than one node are related to solutions of the nonlinear equation with less nodes and different nonlinear eigenvalues;  in section  \ref{sec:le} we consider
the linearized equation for the heterogeneous string and derive its solution; in section \ref{sec:ps} we have considered a string with density $\rho(s) = \gamma (1-s)^\gamma$ and  we have
obtained the threshold values for the occurrence of the nonlinear normal modes using a power series expansion (for the homogeneous 
case, $\gamma=1$, our result reproduces the result of Kolodner \cite{Kolodner55}); finally in section \ref{sec:concl} we state our conclusions and discuss the possible directions of future work.

\section{The heterogeneous helicoseir}
\label{sec:hh}

Consider a string of length $\ell$, that is held fixed at the origin at one end and let $s \ell$ be the distance
from the fixed end along the rope ($0\leq s \leq 1$); we assume that the rope is not homogeneous, with a 
variable density $\rho(s)$.

The mass of the portion of rope from the point $\ell s$ to the free end ($s=1$) is
\begin{equation}
M(s) = \int_s^1 \rho(s) \ell ds \ . \nonumber
\end{equation}
and $M(0) = M$ is the  total mass of the rope.

Now imagine that the rope is rotating around the vertical axis, with constant angular velocity $\omega$;
the vertical and horizontal components of Newton's second law are
\begin{equation}
\begin{split}
\frac{d}{ds} \left( \tau(s) z'(s) \right) &= - \rho(s) g \ell  \\
\frac{d}{ds} \left( \tau(s) r'(s) \right) &= - \rho(s) \omega^2 r(s) \ell  \\
\end{split}
\label{eq_newton}
\end{equation}
where $\tau(s)$ is the tension of the rope at the point $\ell s$.

By integrating the first equation we obtain
\begin{equation}
\tau(s) = \frac{1}{z'(s)} \left[\tau(0) z'(0)  - g \ell^2 \int_0^s \rho(s) ds \right] \nonumber 
\end{equation}
which in turn implies
\begin{equation}
\tau(1) = \frac{1}{z'(1)} \left[\tau(0) z'(0)  - M g \ell \right] \ .  \nonumber 
\end{equation}

Since  the tension vanishes at the free end we deduce 
\begin{equation}
\tau(0) = \frac{M g \ell}{z'(0)}  \nonumber 
\end{equation}
and
\begin{equation}
\tau(s) = \frac{1}{z'(s)} \left[  g \ell \int_s^1 \rho(s) ds \right] = \frac{M(s) g}{z'(s)}  \ .
\label{eq_tau}
\end{equation}

If we express the length of an infinitesimal  piece of string as
\begin{equation}
\ell ds = \sqrt{dr^2 + dz^2}  \nonumber 
\end{equation}
we obtain
\begin{equation}
z'(s) = \sqrt{\ell^2 - r'(s)^2} \ . 
\label{eq_dz}
\end{equation}

Using eqs.(\ref{eq_tau}) and (\ref{eq_dz}) inside the second of eqs.~(\ref{eq_newton}) we 
are left with the differential equation
\begin{equation}
\frac{d}{ds} \left[ \frac{ g \ell^2 \int_s^1 \rho(u) du}{\sqrt{\ell^2 - r'(s)^2}}  r'(s) \right] = - \rho(s) \omega^2 r(s) \ell^2 \ .  \nonumber 
\end{equation}

Before proceeding further in our analysis it is convenient to introduce the dimensionless quantities
\begin{equation}
\begin{split}
R(s) \equiv \frac{r(s)}{\ell} \hspace{0.5cm} , \hspace{0.5cm}
\bar{\rho}(s) \equiv \frac{\ell \rho(s)}{M}  \hspace{0.5cm} , \hspace{0.5cm}
\bar{M}(s) \equiv \frac{M(s)}{M}
\end{split} \nonumber 
\end{equation}
and thus cast the previous equation in the dimensionless form
\begin{equation}
\frac{d}{ds} \left[ \frac{  \bar{M}(s)  R'(s)}{\sqrt{1- R'(s)^2}}  \right] = - \alpha \bar{\rho}(s)  R(s)   \nonumber 
\end{equation}
where $\alpha \equiv \frac{\omega^2 \ell}{g}$.

Equivalently one can write
\begin{equation}
\bar{M}(s) R''(s)  + \bar{M}'(s) \left[ R'(s) - R'(s)^3\right] - \alpha \bar{M}'(s) R(s) (1-R'(s)^2)^{3/2} = 0 \  ,
\label{eq_dif1}
\end{equation}
taking into account that $\bar{\rho}(s) = \bar{M}'(s)$~\footnote{
This equation reduces to equation (9) of Denny~\cite{Denny20} for the case of a homogeneous string 
($\bar{M}(s) = 1 -s$)
\begin{equation}
(1-s) R''(s)  - \left[ R'(s) - R'(s)^3\right] + \alpha  R(s) (1-R'(s)^2)^{3/2} = 0
\label{eq_denny}
\end{equation}
adopting the different convention used in \cite{Denny20} for $s$.}.

If we now  define
\begin{equation}
u(s) \equiv \frac{\bar{M}(s) R'(s)}{\sqrt{1-R'(s)^2}} \nonumber 
\end{equation}
we can cast eq.~(\ref{eq_dif1}) as
\begin{equation}
u''(s) - \alpha \frac{\bar{M}'(s) u(s)}{\sqrt{\bar{M}(s)^2+u(s)^2}} - u'(s) \frac{d\log \bar{M}'(s)}{ds} = 0
\label{eq_dif2}
\end{equation}
that reduces to eq.(14) of ref.~\cite{Kolodner55} for a uniform string.

It is easy to derive the  boundary conditions for this equation; observing that $\bar{M}(1)= 0$  
we have 
\begin{equation}
u(1) \equiv \frac{\bar{M}(1) R'(1)}{\sqrt{1-R'(1)^2}} = 0   \ .
\label{eq_bc1}
\end{equation}

Similarly, observing that $R(0)=0$, we have
\begin{equation}
u'(0) =  \alpha \bar{M}'(0) R(0) = 0 \ .
\label{eq_bc2}
\end{equation}

Equation (\ref{eq_dif2}), subject to the boundary conditions (\ref{eq_bc1}) and (\ref{eq_bc2}), is a {\sl nonlinear eigenvalue equation}, that must be solved numerically. Although we will discuss the numerical solution later on, 
it is worth pointing out that the "shooting method" is the ideal tool to attack this equation (as done for example in
 ref.~\cite{Carter09} for the homogeneous string).

In general it is not possible to find the solutions of eq.~(\ref{eq_dif2}), but we can nonetheless establish some general properties; let $u_{\alpha,\lambda}(s)$  and $u_{\alpha',\lambda'}(s)$ be two different solutions to the eq.~(\ref{eq_dif2}) with 
$u_{\alpha}(0)=\lambda$, such that $u_\alpha(1) = 0$ (and likewise $u_{\alpha'}(0) =\lambda'$  such that $u_{\alpha'}(1)=0$).

By observing that
\begin{equation}
\begin{split}
\mathcal{I} &\equiv \int_{0}^1 \left[ u_{\alpha,\lambda}(s) \frac{d^2 u_{\alpha',\lambda'}(s)}{ds^2} - u_{\alpha',\lambda'}(s) \frac{d^2 u_{\alpha,\lambda}(s)}{ds^2}\right] ds \\
&=  \int_{0}^1  \frac{d}{ds}\left[ u_{\alpha,\lambda}(s) \frac{d u_{\alpha',\lambda'}(s)}{ds} - u_{\alpha',\lambda'}(s) \frac{d u_{\alpha,\lambda}(s)}{ds}\right] ds =  0 
\end{split} \nonumber 
\end{equation}
and  using eq.(\ref{eq_dif2}) we obtain the {\sl nonlinear} orthogonality relation
\begin{equation}
\begin{split}
& \int_{0}^1   \bar{M}'(s) u_{\alpha,\lambda}(s)  u_{\alpha',\lambda'}(s) 
\left( \frac{\alpha'}{\sqrt{\bar{M}(s)^2+u_{\alpha',\lambda'}(s)^2}} - \frac{\alpha}{\sqrt{\bar{M}(s)^2+u_{\alpha,\lambda}(s)^2}}\right) ds  \\
&+ \int_{0}^1  \frac{d\log \bar{M}'(s)}{ds} \left( u_{\alpha,\lambda}(s)  u_{\alpha',\lambda'}'(s)  -u_{\alpha',\lambda'}(s)  u_{\alpha,\lambda}'(s)  \right)  ds  = 0
\end{split}
\label{ortho}
\end{equation}

Apparently this important property has not been detected earlier, even for the case of a homogeneous helicoseir.

There are few important observations to make at this point:
\begin{itemize}
	\item If $u(s)$ is solution to eq.~(\ref{eq_dif2}), then $-u(s)$ is also solution;
	\item eq.~(\ref{ortho}) applies both for $\alpha = \alpha'$ (different modes corresponding to the same angular velocity) 
	and for $\alpha \neq \alpha'$ (different modes corresponding to different angular velocity);
	\item the normalization of the solution is not arbitrary, as  in a linear eigenvalue problem;
	\item the superposition principle does not hold for this problem;
	\item the last term in eq.~(\ref{ortho}) vanishes  for a uniform string ($\bar{M}(s) = 1-s$¸);
\end{itemize}

\section{Conditions of equilibrium}
\label{sec:eq}

Consider an element of the heterogeneous string. There are two forces acting on it: the centrifugal force, acting radially,  and the gravitational force, directed along the vertical axis. 

For an element of mass $dm = \ell \rho(s) ds$ at a distance $\ell s$ from the suspension point (the distance is calculated on the string) these
forces are
\begin{equation}
\begin{split}
d\vec{F}_{centr} =  \ell \rho(s) \omega^2 r(s) ds \ \hat{i}  \hspace{0.5cm} , \hspace{0.5cm}
d\vec{F}_{g} = -\rho(s) g  \ell ds \ \hat{k} \ .
\end{split} \nonumber 
\end{equation}

The torque acting on the infinitesimal mass element is then
\begin{equation}
\begin{split}
d\vec{\tau} &= \rho(s) r(s) \ell \left( \omega^2  z(s) + g \right) ds \ \hat{i} \times \hat{k}
\end{split} \nonumber 
\end{equation}

If we imagine that the string behaves as a straight rigid rod, we may write
\begin{equation}
\begin{split}
r(s) = \ell s \sin\theta \hspace{0.5cm} , \hspace{0.5cm} z(s) = -\ell s \cos\theta
\end{split} \nonumber 
\end{equation}
where  $\theta$ is the angle that the rod forms with the vertical axis.

To achieve the equilibrium, the total torque with respect to the origin must vanish:
\begin{equation}
\tau = l^3 \omega ^2 \cos \theta \ \int_0^1 s^2 \rho (s) \ ds - g l^2  \int_0^1 s \rho (s) \, ds = 0 \ . \nonumber 
\end{equation}

The solution is
\begin{equation}
\cos\theta= \frac{g \int_0^1 s \rho (s)  ds}{l \omega^2 \int_0^1 s^2 \rho (s) \ ds} \ . \nonumber 
\end{equation}

As $|\cos x| \leq 1$, solutions are available only if
\begin{equation}
\alpha \geq \alpha_{\rm threshold} \equiv \frac{\int_0^1 s \rho (s) \ ds}{\int_0^1 s^2 \rho (s) \ ds}\ . \nonumber 
\end{equation}

In particular, for the case of the density $\rho(s) = \gamma (1-s)^{\gamma-1}$ (with $\gamma>0$) one has the condition
\begin{equation}
\alpha \geq \alpha_{\rm threshold}  = 1 + \frac{\gamma}{2} \ .
\label{eq_torque}
\end{equation}

Notice that for $\gamma=1$ we have $\alpha_{\rm threshold} =3/2$, which is the result obtained by Wilson~\cite{Wilson08}.

However this estimate is only approximate because of the assumption that the string behaves as a rigid bar, while moving away from the vertical. Indeed the exact value of $\alpha_{\rm threshold}$ for the homogeneous string calculated by Kolodner, is slightly smaller than the value reported by Wilson (we will later derive the exact expression of $\alpha_{\rm threshold}$ for the special case of density $\rho(s) = \gamma (1-s)^{\gamma-1}$).

Finding an accurate estimate for $\alpha_{\rm threshold}$ would require to know the shape of the string when it departs from the vertical. The approximate value that we have
obtained in eq.~(\ref{eq_torque}) provides a good approximation for a uniform string but it performs poorly for more general cases, where the profile of the string at 
threshold has a significant curvature (a comparison with the exact solution is provided later).

\section{Relating different solutions}
\label{sec:rs}

We have the following theorem:

\begin{theorem}[Matrioska Effect]
Consider a heterogeneous string with density $\rho(s)$,  $0 \leq s \leq 1$. Let $u(s)$ be a solution of eq.~(\ref{eq_dif2})
with $n$ nodes and $n$ extrema (maxima and minima).
We call $s^{(0)}_k$ and $s^{(e)}_k$ , with $k=1,\dots,n$, the positions at which the nodes and the extrema are located
(clearly $s^{(e)}_1 = 0$ and $s^{(0)}_n = 1$  and $s^{(e)}_1 < s^{(0)}_1 < s^{(e)}_2 < \dots < s^{(e)}_{n} < s^{(0)}_n$).     

On any subinterval $(s^{(e)}_k, s^{(0)}_j)$, if $s_k^{(e)} < s_j^{(0)}$, or  $(s^{(0)}_j,s^{(e)}_k$, if $s_k^{(e)}) > s_j^{(0)}$,
the function
\begin{equation}
G(\eta) \equiv \frac{u(s(\eta))}{|s^{(e)}_k-s^{(0)}_j|} \hspace{0.5cm} , \hspace{0.5cm} s(\eta) = s^{(e)}_k + \eta (1-s_j^{(e)}) \ ,
\end{equation}
with $0 \leq \eta \leq 1$,  is a solution of the equation
\begin{equation}
{\frac{d^2G(\eta)}{d\eta^2}} - \left[ \frac{\alpha (1-s_k^{(e)})}{\sqrt{\mathcal{M}^2(\eta) +G^2(\eta)}} {\tfrac{d\mathcal{M}(\eta)}{d\eta}} G(\eta) + 
{\frac{\tfrac{d^2\mathcal{M}(\eta)}{d\eta^2}}{\tfrac{d\mathcal{M}(\eta)}{d\eta}} \frac{dG(\eta)}{d\eta}}\right] = 0
\end{equation}
with $\mathcal{M}(\eta) \equiv \frac{\bar{M}(s(\eta))}{|s_k^{(e)}-s_j^{(0)}|}$ and boundary conditions $G'(0) = 0$ and $G(1)=0$.
\end{theorem}

{\bf Proof:} To start with we consider a homogeneous string ($\bar{M}(s)=1-s$) and a solution to eq.~(\ref{eq_dif2}) with $n$ nodes.

We introduce the rescaled variable
\begin{equation}
\sigma \equiv \frac{s-s^{(e)}_k}{|s^{(0)}_j-s^{(e)}_k|}  \ , \hspace{1cm} k,j =1,2, \dots,n
\label{eq_sigma}
\end{equation} 
and express $s$ in terms of $\sigma$
\begin{equation}
s(\sigma) = s^{(e)}_k + \sigma |s^{(e)}_k - s^{(0)}_j | \ ,
\label{eq_s}
\end{equation}
with $0 \leq \sigma \leq L$ and $L \equiv \frac{(1-s_k^{(e)})}{|s^{(e)}_k - s^{(0)}_j |}$.

We then define
\begin{equation}
\varUpsilon(\sigma) \equiv  u(s(\sigma))/A \ ,
\label{eq_ups}
\end{equation}
where $A$ is a constant to be determined; eq.~(\ref{eq_dif2}) takes the form
\begin{equation}
\frac{d^2\varUpsilon}{d\sigma^2} + \frac{\alpha (s^{(e)}_k - s^{(0)}_j )^2 \  \varUpsilon(\sigma) }{ { A \sqrt{\frac{|s^{(e)}_k - s^{(0)}_j |^2}{A^2} \left( L -\sigma \right)^2  + \varUpsilon^2(\sigma) }}} =0 \ .  \nonumber
\end{equation}

If we choose $A= |s^{(e)}_k - s^{(0)}_j|$ and define
\begin{equation}
\begin{split}
\tilde{\alpha} \equiv \alpha \ |s^{(e)}_k - s^{(0)}_j | \hspace{0.6cm} , \hspace{0.6cm}
\tilde{M}(\sigma) \equiv  L -\sigma \\
\label{eq_rescaled}
\end{split}
\end{equation}
we have that $\varUpsilon(\sigma)$ obeys the differential equation
\begin{equation}
\frac{d^2\varUpsilon}{d\sigma^2} -  \frac{\tilde{\alpha} \ {\tfrac{d\tilde{M}(\sigma)}{d\sigma}} \ \varUpsilon(\sigma) }{  \sqrt{\tilde{M}(\sigma)^2 + \varUpsilon^2(\sigma) }} =0 \ ,  \nonumber
\end{equation}
which corresponds to the eq.~(\ref{eq_dif2}) with boundary conditions
$\left. \frac{d\varUpsilon}{d\sigma} \right|_{\sigma=0} =   0$,  $\varUpsilon(L)     = 0$.

We conclude that $\varUpsilon(\sigma)$ is the solution for a homogeneous helicoseir of length $L$ and total mass $L$ (we are working with  dimensionless quantities). 

We can perform a last rescaling; define
\begin{equation}
\begin{split}
\eta \equiv \frac{\sigma}{L} \  , \ G(\eta) \equiv \Upsilon(\eta L) \ ,  \  \mathcal{M}(\eta) \equiv \tilde{M}(\eta L) = L (1-\eta)
\end{split} \label{eq_last_rescaling}
\end{equation}
and obtain that $G(\eta)$ is a solution of the differential equation
\begin{equation}
\frac{d^2G}{d\eta^2} - \frac{\dbtilde{\alpha}  \ {\tfrac{d\mathcal{M}(\eta)}{d\eta}} G(\eta)}{\sqrt{\mathcal{M}^2(\eta)+G^2(\eta)}} = 0 
\end{equation}
where $\dbtilde{\alpha} \equiv  \tilde{\alpha} L  = \alpha (1-s_k^{(e)})$. This equation describes a homogeneous helicoseir of unit length and total mass $L$.
Notice that for the special case $s_j^{(0)}=1$, we have $L=1$ and this is the original equation, eq.~(\ref{eq_dif2}), with $\dbtilde{\alpha} < \alpha$.

Let us now consider the general case of an inhomogeneous string with density $\rho(s)$. We use eqs.(\ref{eq_s}) and (\ref{eq_ups}) and we
find that $\varUpsilon(\sigma)$ obeys the equation
\begin{equation}
\begin{split}
\frac{d^2\varUpsilon}{d\sigma^2} - \left[ \frac{\alpha |s^{(e)}_k -s^{(0)}_j|^2 \ \frac{d\bar{M}}{ds}}{\sqrt{\bar{M}(s(\sigma))^2 + A^2 \varUpsilon(\sigma)^2}} \ \varUpsilon(\sigma) + \frac{|s^{(e)}_k -s^{(0)}_j| \  \frac{d^2\bar{M}}{ds^2}}{ \frac{d\bar{M}}{ds}}  \ \frac{d\varUpsilon}{d\sigma}\right] = 0
\end{split} \nonumber
\end{equation}

We choose $A=  |s^{(e)}_k -s^{(0)}_j|$ and $\tilde{M}(\sigma) \equiv \frac{\bar{M}(s(\sigma))}{|s^{(e)}_k -s^{(0)}_j|}$, which implies:
\begin{equation}
\begin{split}
\frac{d\bar{M}}{ds}     &= \frac{d\bar{M}}{d\sigma} \ \frac{d\sigma}{ds} = \frac{d\tilde{M}}{d\sigma} \\
\frac{d^2\bar{M}}{ds^2} &= \frac{d^2\bar{M}}{d\sigma^2} \ \left(\frac{d\sigma}{ds}\right)^2 =  \frac{1}{|s^{(e)}_k -s^{(0)}_j|} \frac{d^2\tilde{M}}{d\sigma^2} \\ 
\end{split} \nonumber
\end{equation}

Therefore the equation above reduces to 
\begin{equation}
\begin{split}
\frac{d^2\varUpsilon}{d\sigma^2} - 
\left[ \frac{\alpha |s^{(e)}_k -s^{(0)}_j| \ \frac{d\tilde{M}}{d\sigma}}{\sqrt{\tilde{M}(\sigma)^2 + \varUpsilon(\sigma)^2}} \ \varUpsilon(\sigma) + \frac{\frac{d^2\tilde{M}}{d\sigma^2}}{ \frac{d\tilde{M}}{d\sigma}}  \ \frac{d\varUpsilon}{d\sigma}\right] = 0
\end{split}
\label{eq_helico}
\end{equation}
with boundary conditions $\left. \frac{d\varUpsilon}{d\sigma} \right|_{\sigma=0} =  0$ and $\varUpsilon(L) =  0$. 

Eq.~(\ref{eq_helico}) has the same form of the original equation, with $\alpha \rightarrow \tilde{\alpha} \equiv \alpha |s^{(e)}_k -s^{(0)}_j|$ and with length $L$. 

We can perform a last rescaling, described in eqs.~(\ref{eq_last_rescaling}) and we find that $G(\eta)$ is a solution to the heterogeneous helicoseir problem with 
{$\alpha \rightarrow \tilde{\tilde{\alpha}}\equiv\alpha L |s_k^{(e)} - s_j^{(0)}|$} and {$\tilde{M}(\sigma) \rightarrow \mathcal{M}(\eta)$}:
\begin{equation}
{\frac{d^2G(\eta)}{d\eta^2}} - \left[ \frac{\tilde{\tilde{\alpha}}}{\sqrt{\mathcal{M}^2(\eta) +G^2(\eta)}} {\tfrac{d\mathcal{M}(\eta)}{d\eta}} G(\eta) + 
{\frac{\tfrac{d^2\mathcal{M}(\eta)}{d\eta^2}}{\tfrac{d\mathcal{M}(\eta)}{d\eta}} \frac{dG(\eta)}{d\eta}}\right] = 0 \nonumber
\end{equation}
{That in terms of our original $\alpha$ we finally obtain}
\begin{equation}
{\frac{d^2G(\eta)}{d\eta^2}} - \left[ \frac{\alpha (1-s_k^{(e)})}{\sqrt{\mathcal{M}^2(\eta) +G^2(\eta)}} {\tfrac{d\mathcal{M}(\eta)}{d\eta}} G(\eta) + 
{\frac{\tfrac{d^2\mathcal{M}(\eta)}{d\eta^2}}{\tfrac{d\mathcal{M}(\eta)}{d\eta}} \frac{dG(\eta)}{d\eta}}\right] = 0
\end{equation}
with boundary conditions $G'(0)=0$ and $G(1)=0$. This proves the theorem.

In particular for the case 
\begin{equation}
\bar{M}(s) = (1-s)^\gamma  \nonumber
\end{equation}
we have
\begin{equation}
{\mathcal{M}(\eta) = (1-s_k^{(e)})^{\gamma} (1-\eta)^\gamma}
\end{equation}

The total mass of this new string is then $\mathcal{M}(0) =  (1-s_k^{(e)})^{\gamma}$.

\section{Linearized equations}
\label{sec:le}

Following Kolodner \cite{Kolodner55}, we set $u(s) = \epsilon \tilde{u}(s)$, with $\epsilon\rightarrow 0$, and substitute in the nonlinear eigenvalue equation. To leading order we obtain the equation
\begin{equation}
\tilde{u}''(s) - \alpha \frac{M'(s)}{|M(s)|} \tilde{u}(s) - \frac{M''(s)}{M'(s)} \tilde{u}'(s) = 0 \nonumber
\end{equation}
with $\tilde{u}'(0)=0$ and $\tilde{u}(1) =0$.

For the special case of a uniform rope  this equation reduces to 
\begin{equation}
\tilde{u}''(s) + \alpha \frac{1}{1-s} \tilde{u}(s)  = 0 \nonumber
\end{equation}
with solution
\begin{equation}
\tilde{u}(s) = C \sqrt{\alpha (1-s)} \  J_1\left(2 \sqrt{\alpha (1-s) }\right) \nonumber
\end{equation}
for $\alpha_n = \frac{\sigma_{0,n}^2}{4}$, $\sigma_{k,n}$ being the $n^{th}$ zero of the Bessel function $J_k(x)$, $k=0,1,2,\dots$~\cite{Kolodner55}. $C$ is a normalization constant.

Let us consider the more general case 
\begin{equation}
M(s) = (1-s)^\gamma \nonumber
\end{equation}
where $\gamma \geq 1$. 

The linearized equation in this case is 
\begin{equation}
\tilde{u}''(s)  - \left(\frac{1-\gamma }{1-s}\right) \ \tilde{u}'(s) + \frac{\alpha  \gamma }{1-s} \tilde{u}(s)=0 \nonumber
\end{equation}
and its solution has the form
\begin{equation}
\tilde{u}(s) = C (\alpha  (1-s))^{\gamma /2} J_{\gamma }\left( 2\sqrt{\gamma \alpha (1-s) }\right) \nonumber
\end{equation}

For $\gamma$ integer it is straightforward to see that  admissible solutions occur only for
\begin{equation}
\alpha = \frac{\sigma_{\gamma-1,n}^2}{4 \gamma} \label{eq_egv}
\end{equation}
which reduces to the results of \cite{Kolodner55} for $\gamma=1$.

As for the case of a homogeneous string, the spectrum of the linearized equation is discrete with eigenvalues (\ref{eq_egv}),
which are the thresholds for the appearance of new modes for the nonlinear equation.

\section{Power series expansion}
\label{sec:ps}

Let us consider the case
\begin{equation}
\bar{M}(s) = (1-s)^\gamma \nonumber
\end{equation}
which includes the homogeneous string as a  special case, $\gamma=1$.

A simple analysis shows that, for $\gamma \geq 1$, the solution should behave as $u(s) \approx (1-s)^\gamma$,
for $s\rightarrow 1^-$. Therefore we write 
\begin{equation}
u(s)= (1-s)^\gamma v(s) \nonumber
\end{equation}

After substituting inside eq.~(\ref{eq_dif2}) we obtain the equation
\begin{equation}
(1-s) v''(s)-(\gamma +1) v'(s)+\frac{\alpha  \gamma v(s)}{\sqrt{1+v(s)^2}} = 0  \nonumber
\end{equation}

For $s \rightarrow 1^-$ we can represent $v(s)$ as a power--series expansion about $s=1$:
\begin{equation}
v(s) = \sum_{k=0}^\infty c_k (1-s)^k \nonumber
\end{equation}
and obtain the coefficients $c_k$ of the series by sustituting  the series in the differential equation;
the first few coefficients, expressed in terms of $c_0$, read
\begin{equation}
\begin{split}
c_1 &= -\frac{\alpha  \gamma  c_0}{(\gamma +1) \sqrt{c_0^2+1}} \\
c_2 &= \frac{\alpha ^2 \gamma ^2 c_0}{2 \left(\gamma ^2+3 \gamma +2\right) \left(c_0^2+1\right)^2} \\
c_3 &= \frac{\alpha ^3 \gamma ^3 c_0 \left(3 (\gamma +2) c_0^2-\gamma -1\right)}{6 (\gamma +1)^2 (\gamma +2) (\gamma +3) \left(c_0^2+1\right)^{7/2}} \\
c_4 &= \frac{\alpha ^4 \gamma ^4 c_0 \left(12 (\gamma +2) (\gamma +3) c_0^4-3 (5
	\gamma  (\gamma +4)+17) c_0^2+(\gamma +1)^2\right)}{24 (\gamma +1)^3
	(\gamma +2) (\gamma +3) (\gamma +4) \left(c_0^2+1\right)^5} \\
\dots &
\end{split} \nonumber
\end{equation}
We have calculated exactly the coefficients up to $c_{13}$.

By introducing the variable $\phi \equiv \alpha \gamma$, $v(s)$ takes a simpler form
\begin{equation}
\begin{split}
v(s) &= \frac{c_0 (s-1) \phi }{(\gamma +1) \sqrt{c_0^2+1}} + \frac{c_0 (s-1)^2 \phi ^2}{2 \left(\gamma ^2+3 \gamma +2\right) \left(c_0^2+1\right){}^2} \\
&+ \frac{c_0 (s-1)^3 \phi ^3 \left(-3 (\gamma +2) c_0^2+\gamma +1\right)}{6 (\gamma +1)^2 (\gamma +2) (\gamma
   +3) \left(c_0^2+1\right){}^{7/2}} + \dots
\end{split}
\end{equation}
and we can consider $F(c_0,\phi) \equiv u'(0)$:
\begin{equation}
F(c_0,\phi) = \frac{c_0 \phi }{\sqrt{c_0^2+1}} -\frac{c_0 \phi ^2}{2 (\gamma +1) \left(c_0^2+1\right){}^2}
-\frac{c_0 \phi ^3 \left(3 (\gamma +2) c_0^2-\gamma -1\right)}{6 (\gamma +1)^2 (\gamma +2)
   \left(c_0^2+1\right){}^{7/2}} + \dots
\end{equation}

Clearly $c_0=0$ corresponds to the trivial solution, but as $\alpha$ increases, $\left. \frac{\partial F}{\partial c_0} \right|_{c_0=0}$ may suddenly change sign. When this occurs, it means that one has reached the critical value of $\alpha$ where a non--trivial  solution suddenly appears.

The explicit expression for $\left. \frac{\partial F}{\partial c_0} \right|_{c_0=0}$ that we obtain is
\begin{equation}
\begin{split}
\left. \frac{\partial F}{\partial c_0} \right|_{c_0=0} &= -\gamma+\phi -\frac{\phi ^2}{2+2 \gamma }+\frac{\phi ^3}{6 (1+\gamma ) (2+\gamma)}-\frac{\phi ^4}{24 (1+\gamma ) (2+\gamma ) (3+\gamma )} \\
&+ \frac{\phi^5}{120 (1+\gamma ) (2+\gamma ) (3+\gamma ) (4+\gamma )} + \dots
\end{split} \nonumber 
\end{equation}

We can clearly guess the full series to be
\begin{equation}
\begin{split}
\left. \frac{\partial F}{\partial c_0} \right|_{c_0=0} &= \sum_{k=0}^\infty \frac{(-1)^{k+1} }{k!} \frac{ \Gamma (\gamma +1) }{\Gamma (\gamma+k )} \  \phi^k = -\gamma  \ _0F_1(;\gamma ;-\phi )
\end{split} \nonumber
\end{equation}
where $_0F_1$ is the hypergeometric function.

For integer values of $\gamma$ this function vanishes precisely at 
\begin{equation}
\alpha = \frac{\sigma_{\gamma-1,n}^2}{4 \gamma}  \nonumber
\end{equation}
where $\sigma_{\gamma-1,n}$ it the $n^{th}$ zero of the Bessel function of order $\gamma-1$ (it is easy to check that  
our expression reduces to the expression found by Kolodner for $\gamma=1$, as it should).

\begin{figure}
\begin{center}
\bigskip\bigskip\bigskip
\includegraphics[width=9cm]{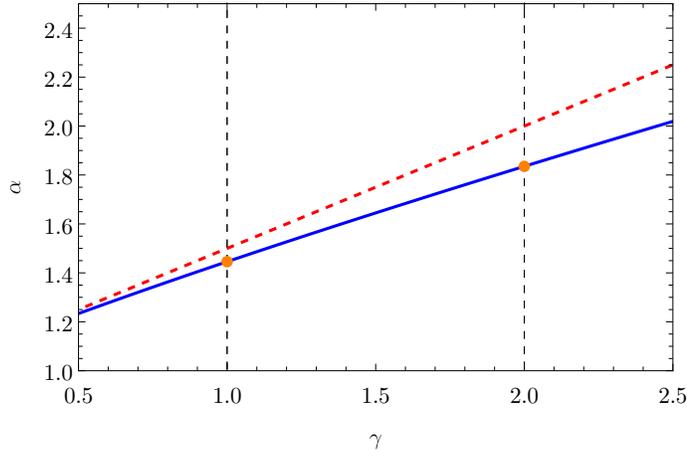}
\bigskip
\caption{Solution of $\left. \frac{dF}{dc_0} \right|_{c_0=0}=0$ (blue curve) compared to the threshold estimated in eq.~(\ref{eq_torque}) (red dashed curve). The orange dots for integer values of $\gamma$ correspond to the the values $\alpha = \frac{\sigma_{\gamma-1,n}^2}{4 \gamma}$.}
\label{Fig_1}
\end{center}
\end{figure}

In Fig.~\ref{Fig_1} we plot the lowest root of $\left. \frac{dF}{dc_0} \right|_{c_0=0}=0$ as a function of $\gamma$. For $\gamma=1$ 
eq.~(\ref{eq_torque}) slightly overstimates the true root. 

\begin{figure}
\begin{center}
\bigskip\bigskip\bigskip
\includegraphics[width=7cm]{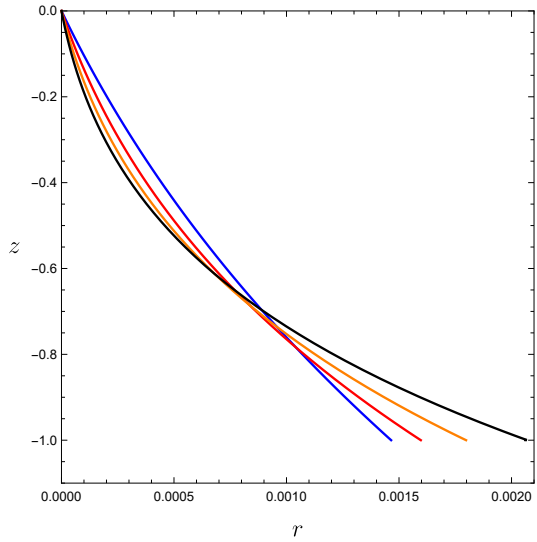}
\bigskip
\caption{Profile of the string for $\gamma = 1$ (blue), $2$ (red), $3$ (orange) and $4$ (black) for $\alpha = \frac{\sigma_{\gamma-1,n}^2}{4\gamma} +10^{-6}$. Notice that the horizontal and vertical axes are not in scale.}
\label{Fig_2}
\end{center}
\end{figure}

In Fig.~\ref{Fig_2} we plot the profile of the string, for four different densities and for $\alpha = \frac{\sigma_{\gamma-1,n}^2}{4\gamma} +10^{-6}$ (very close to the critical value). Clearly, as $\gamma$ is increased, the curvature of the profile increases as well and the
approximation as a straight line becomes less and less accurate.

\begin{figure} 
\begin{center}
\bigskip\bigskip\bigskip
\includegraphics[width=6cm]{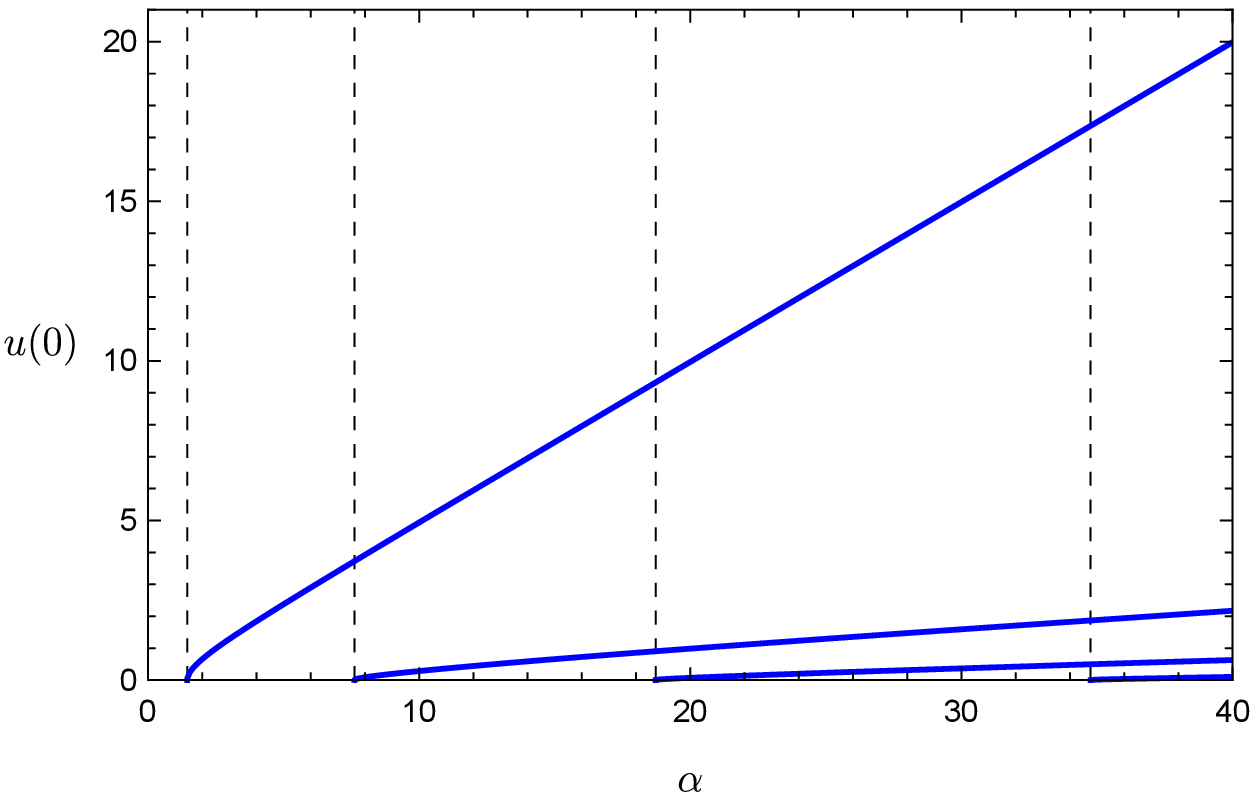} \hspace{1cm}
\includegraphics[width=6cm]{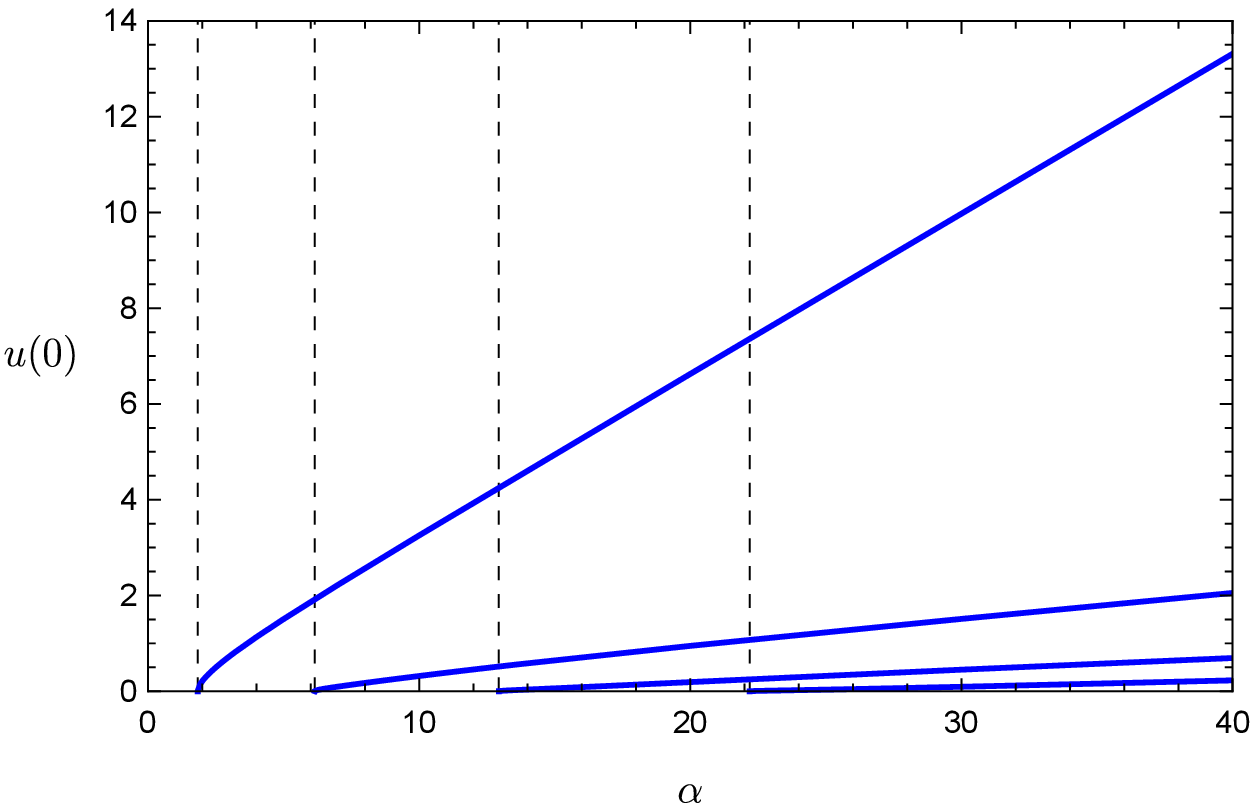}
\caption{{$u(0)$ as a function of $\alpha$ for $\gamma=1$ (left plot) and $\gamma=2$ (right plot). The vertical lines are the eigenvalues of the linearized equation $\frac{\sigma_{\gamma-1,n}^2}{4 \gamma}$. }}
\label{Fig_alphagamma}
\end{center}
\end{figure}

In Fig.~\ref{Fig_alphagamma} we plot $u(0)$ for $\gamma=1$ (left plot) and $\gamma=2$ (right plot) as a function of $\alpha$: the vertical lines are the eigenvalues of the linearized equations in eq.~(\ref{eq_egv}), at which the modes of the nonlinear equation 
start appearing. For $\alpha \gg 1$, it is seen that $u(0)$ grows linearly with $\alpha$ and in particular we have found that 
the fundamental solutions for $\gamma=1$ and $\gamma=2$, respectively tend to 
\begin{equation}
\begin{split}
u(s) \approx \frac{\alpha}{2} (1-s^2) \hspace{1cm} &, \hspace{1cm} \gamma=1 \ , \ \alpha \gg 1 \\
u(s) \approx \frac{\alpha}{3} (1-s)^2 (1+2s) \hspace{1cm} &, \hspace{1cm} \gamma=2 \ , \ \alpha \gg 1 \\
\end{split}
\nonumber
\end{equation}

These results clearly suggest
\begin{equation}
\begin{split}
u(s) \approx \frac{\alpha}{\gamma+1} (1-s)^\gamma (1+\gamma s) \hspace{1cm} &, \hspace{1cm} \alpha \gg 1 \\
\end{split}
\label{eq_alpha}
\end{equation}

Indeed it is easy to see that eq.~(\ref{eq_alpha}) is a solution of eq.~(\ref{eq_dif2}) to order $\alpha$, for $\alpha \rightarrow \infty$.

We will now use the power series expansion of the solution about $s=1$ to estimate the behavior of the amplitude $u(0)$ as a function of $\alpha$:
clearly the leading coefficient $c_0$ is a function of $\alpha$ and the numerical solutions plotted in Fig.~\ref{Fig_alphagamma} suggest that it grows
linearly in $\alpha$ for $\alpha \gg 1$. 

This finding is confirmed if we write $c_0 = \alpha \bar{c}_0$, where $\bar{c}_0$ is just a number and substitute it in the power series expansion of $u(s)$;
in this case we find that
\begin{equation}
u'(0) = O\left(\frac{1}{\alpha}\right)
\end{equation}
and the boundary condition is automatically satisfied at $s=0$. 

Correspondingly we find that the amplitude is given by the expression
\begin{equation}
u(0) = \frac{\alpha}{1+\gamma} +  O\left(\frac{1}{\alpha}\right)
\end{equation}
which agrees with eq.~(\ref{eq_alpha}).

As an example, the numerical solution for $\gamma =2$ and $\alpha = 100$ has $u(0) \approx 33.325411$, that is remarkably close to the value predicted by the asymptotic formula, 
$u^{(asym)}(0) = 100/3 = 33.3333\bar{3}$.

\section{Conclusions}
\label{sec:concl}

We have discussed the case of a inhomogeneous helicoseir, i.e. of a inhomogeneous string that is set to rotate about a vertical axis, in a
constant gravitational field. The string can assume configurations in which the shape is not changing with time, but it rotates rigidly
about the vertical axis. The case of a homogeneous string was studied long time ago by Wilson~\cite{Wilson08} and later by Kolodner~\cite{Kolodner55}. We have derived the nonlinear eigenvalue equation for the case of an arbitrary density and we have established the orthogonality conditions obeyed by the solutions. For the special case of an inhomogeneous string with 
$M(s) = (1-s)^\gamma$ (which also includes the homogeneous case) we have been able to find the critical values of angular velocity where
new modes of the nonlinear eigenvalue equation appear, thus generalizing the analysis of Kolodner. These critical values correspond to the eigenvalues of the corresponding linearized eigenvalue equation, precisely as for the homogeneous case.
Finally, we have found the interesting property, that the "dissection" of a solution of the nonlinear eigenvalue equation with $n > 1$ nodes, corresponds to a solution of a nonlinear equation with different eigenvalue (jokingly, we refer to this occurrence
as a "matrioska effect").

\section*{Acknowledgements}
The research of P.A. was supported by Sistema Nacional de Investigadores (M\'exico). 
The plots in this paper have been plotted using {\rm MaTeX} \cite{szhorvat}.

\end{document}